\documentclass[twocolumn,showpacs,aps]{revtex4}
\usepackage[dvips]{graphicx}

\usepackage{times}

\begin{document}

\title{Invertible condition of quantum Fisher information matrix for a mixed qubit}

\author
{Ping Yue $^{1,~2}$, Li Ge $^{3}$, and
Qiang Zheng $^{1,~3}$,
\footnote{ Corresponding author: qz@csrc.ac.cn}
}
\address{
$^{1}$ School of Mathematics and Computer Science, Guizhou Normal
University, Guiyang, 550001, China
\\
$^{2}$ School of Biology and Engineering, Guizhou Medical
University, Guiyang, 550001, China
\\
$^{3}$ Division of Quantum Physics and Quantum Information, Beijing Computational Science Research Center, Beijing 100084, China
}

\begin{abstract}
Estimating multiparamter simultaneously as precise as possible is an important goal of
quantum metrology. As a first step to this end, here we give a condition determining whether two arbitrary parameters can be estimated simultaneously for a qubit in the mixed state. An application of this condition
is shown.
\end{abstract}


\maketitle

\section{Introduction}

Quantum metrology~\cite{VGio11}, which exploits the quantum resource to significantly
enhance the sensitivity, is an active field in recent years.
It has wide application in enhancing time standard~\cite{Udem02, Bollinger}, probing gravitational waves~\cite{Aasi2}, and magnetometry~\cite{Taylor08}. Most problems in quantum metrology can be casted
into the parameter estimation. In the single parameter case, according to quantum Cram\'{e}r-Rao inequality,
quantum Fisher information (QFI) plays a central role \cite{Escher11, Guta12, zyou14, jin13, qz14, wzhong13}. The QFI is also closely connected with other quantities, especially entanglement~\cite{Smerzi09}, non-Markovianity~\cite{xmLu10}, and spin squeezing~\cite{jma11}.

For most system, the property is usually related to the multiple parameters.
In this case, it is reasonable to only estimate one
parameter by preparing the independently probe state in each run. However, this
scheme, which needs a quite a lot of resource, is
not an optimal one to estimate the multiple parameters,
as one should adjust a different experimental setup for each parameter. Principally,
this drawback can be overcome by estimating the parameters simultaneously, and the
problem changes to multiparameter estimation
\cite{yuen73, Helstrom74}.
In this case, quantum Fisher information matrix (QFIM)
\cite{Ballester04, Walmsley14} plays the equivalently key role as the QFI
in the single parameter estimation.

In this paper, we theoretically study the quantum multiparameter estimation making use of a qubit,
as it is a well-known basic element in the field of quantum information and quantum metrology.
Our study is also
motivated by recent experimental achievement on a qubit \cite{Vijay12}.
The main result of this paper is that based on the invertibility of the QFIM,
a condition is given to determine whether two (arbitrary) parameters
can be estimated simultaneously for a qubit in the mixed state.
We also investigate the QFIM of a dissipative
qubit as an example for applying the invertible condition.

This paper is organized as follows.
The basic properties of the QFI and the QFIM are reviewed in Sec.~II. In Sec.~III, we give a condition determining whether two arbitrary parameters
can be estimated simultaneously for a qubit in the mixed state. We also show an example as
an application of this condition. Finally, a summary is provided in the last section.

\section{QFI and QFIM}
In this section, we will review the main aspects of QFI and QFIM. Let $\varphi$ denote a single parameter to be estimated, and $p_{i}(\varphi)$ be the probability density with measurement outcome $\{ x_{i} \}$ for a discrete observable $X$ conditioned on the fixed parameter $\varphi$. The classical Cram\'{e}r-Rao inequality~\cite{Holevo} gives the bound of the variance $\mathrm{Var}(\hat{\varphi})$
for an unbiased estimator $\hat{\varphi}$
\begin{equation}
\begin{array}{llll}
\mathrm{Var}(\hat{\varphi}) \geq \frac{1}{H_{\varphi}},
\end{array}
\end{equation}
where the classical Fisher information is defined as \cite{Fisher}
$H_{\varphi}= \sum_{i}p_{i}(\varphi)[\frac{\partial}{\partial \varphi} \ln p_{i}(\varphi)]^2$.

Extending to quantum regime, in order to determine the ultimate bound to precision posed by quantum mechanics, the Fisher information must be maximized over all possible measurements~\cite{paris09}. By introducing the symmetric logarithmic derivative $L_{\varphi}$, which is determined by
\begin{equation}
\begin{array}{llll}
\frac{ \partial}{ \partial \varphi }\rho_{\varphi}=\frac{1}{2} (\rho_{\varphi}
L_{\varphi}+ L_{\varphi} \rho_{\varphi} ),
\end{array}
\end{equation}
the so-called quantum Cram\'{e}r-Rao inequality gives a bound to the variance of any unbiased estimator~\cite{Caves94}:
\begin{equation}
\begin{array}{llll}
\mathrm{Var}(\hat{\varphi}) \geq \frac{1}{H_{\varphi}} \geq \frac{1}{F_{\varphi}}.
\end{array}\label{crineqB}
\end{equation}
Here, the QFI of a quantum state $\rho_{\varphi}$ with respect to the parameter $\varphi$ is defined as~\cite{Caves96}
\begin{equation}
\begin{array}{llll}
F_{\varphi}= \mathrm{Tr} ( \rho_{\varphi} L_{\varphi}^2 ).
\end{array}
\end{equation}

Moreover, the QFI is also related to the Bures distance through
\begin{equation}
D_{B}^2[\rho_{\varphi}, \rho_{\varphi+d\varphi}]= \frac{1}{4}F_{\varphi}d \varphi^2,
\end{equation}
where the Bures distance is defined as~\cite{Nielsen00} $ D_{B}[\rho, \sigma]=[ 2(1-\mathrm{Tr}\sqrt{\rho^{1/2} \sigma \rho^{1/2}})]^{1/2}$, which measures the distance between two quantum states $\rho$ and $\sigma$.

In the scenario that a set of parameters $\Phi=\{ \varphi_{j} \}$ is involved, the QFI is replaced by the QFIM~\cite{Genoni08, Whaley09}, which is defined as
\begin{equation}
\begin{array}{llll}
\textbf{F}_{\Phi, ij}= \frac{1}{2}\mathrm{Tr}( \rho_{\Phi} \{L_{i}, L_{j}\} ).
\end{array}
\end{equation}
Here $L_{i}$ is the symmetric logarithmic derivative corresponding to the parameter $\varphi_{i}$, and $\{A, B\}=AB+BA$ is the anticommutator. It is apparent that $\textbf{F}_{\Phi}$ is a real symmetric matrix and its diagonal element should reduce to the single-parameter QFI.

With the spectrum decomposition $\rho_{\varphi}= \sum_{k} \lambda_{k}|k\rangle \langle k|$, the element of the QFIM can be explicitly expressed as
\begin{equation}
\textbf{F}_{\Phi, ij}= \textbf{F}^{C}_{\Phi, ij}+\textbf{F}^{Q}_{\Phi, ij}
\end{equation}
with the classical Fisher information matrix
\begin{eqnarray}
\textbf{F}^{C}_{\Phi, ij}&=&\sum_{k, \lambda_{k}>0} \frac{(\partial_{i}\lambda_{k}) (\partial_{j}\lambda_{k})}{\lambda_{k}},
\end{eqnarray}
and the quantum part of the information matrix
(the summation only for the terms with $\lambda_{k}+\lambda_{k'}>0$)
\begin{eqnarray}
\textbf{F}^{Q}_{\Phi, ij}=&&\sum_{k, k'} \frac{(\lambda_{k}-\lambda_{k'})^2}{\lambda_{k}+\lambda_{k'}}\times \nonumber\\
&&(\langle k | \partial_{i}k' \rangle \langle\partial_{j}k'| k\rangle
\,+ \langle k | \partial_{j}k' \rangle \langle\partial_{i}k'| k\rangle).
\end{eqnarray}
Here $| \partial_{i}k\rangle$ represents the partial derivative of $| k\rangle$ with respect to $\varphi_{i}$.

The precision of the estimation of $\Phi$ is governed by the covariance matrix of parameters
$\Theta_{ij}=\langle \varphi_{i} \varphi_{j} \rangle- \langle \varphi_{i}\rangle \langle \varphi_{j} \rangle$.
The quantum Cram\'{e}r-Rao bound of single parameter transfers to matrix inequality
\cite{Monras10, Kim13}
\begin{equation}
\begin{array}{llll}
\Theta \geq \textbf{F}_{\Phi} ^{-1}.
\end{array} \label{mCRB}
\end{equation}
Generally speaking, this bound cannot be achieved. The attainability of
quantum Cram\'{e}r-Rao bound Eq.~(\ref{mCRB}) has been extensively discussed recently \cite{Fujiwara01}.
Taking the trace of two sides of Eq.~(\ref{mCRB}), a lower bound on the total variance of all the estimated parameters follows as ~\cite{Walmsley13, hfan14}
\begin{equation}
\begin{array}{llll}
|\Delta \Phi|^2 \equiv \mathrm{Tr}[\Theta]  \geq \mathrm{Tr}[\textbf{F}_{\Phi} ^{-1}].
\end{array} \label{jointCR}
\end{equation}
According to Eq.~(\ref{jointCR}), if the inverse matrix of $\textbf{F}_{\Phi}$ does not exist, that is $\mathrm{Det}(\textbf{F}_{\Phi})= 0$, the error of the joint estimation of the parameters $\Phi$ is unbounded.
It should be noted that the invertibility of the QFIM is only a
necessary condition for joint estimability of the corresponding parameters.

\section{Invertible condition of QFIM for a qubit}
As shown by Eq.~(\ref{jointCR}),
the lower bound on the total variance of all the estimated parameters
is related to the inverse matrix of QFIM. Thus, the invertibility is an basic
property of the QFIM. In the section, we give a condition to determine whether an inverse matrix of QFIM for a
mixed qubit state with two arbitrary parameters exists. Denoting the two arbitrary parameters as $x$ and $y$ (for simplicity, we assume they are real numbers), the mixed state of the qubit can be generally expressed as
\begin{equation}
\begin{array}{llll}
\rho=\lambda(x, y) |\phi_{1}\rangle \langle \phi_{1}|+(1-\lambda(x, y)) |\phi_{2}\rangle \langle \phi_{2}|
\end{array}
\end{equation}
with the constraints $0 <\lambda (x, y)<1$ and $\langle \phi_{i}|\phi_{j}\rangle=\delta_{ij}$. Thus, the eigenvectors of $\rho$ can be generally expressed as
\begin{equation}
\begin{array}{llll}
|\phi_{1}\rangle=[\cos(h(x, y)) \exp(i\theta(x, y)), \sin(h(x, y))],
\\
\\
|\phi_{2}\rangle=[-\sin(h(x, y)), \cos(h(x, y))\exp(-i\theta(x, y))].
\end{array}
\end{equation}
Here $h(x, y)$ and $\lambda(x, y)$ are two arbitrary real functions.

In this paper, we only consider a simple case $\theta(x, y)\equiv \theta_{0}$, where
$\theta_{0}$ is an arbitrary real function independent of $x$ and $y$.
After some straightforward calculations, the QFIM of $\rho$ in terms of $x$ and $y$ is
obtained as
\begin{equation}
\textbf{F}_{x,~y}=\left(                 
\begin{array}{cc}   
 F_{11} & F_{12} \\  
 F_{21} &  F_{22} \\  
 \end{array}
 \right) \label{qfisMD}
\end{equation}
with the elements
\begin{equation}
\begin{array}{llll}
F_{11}=\frac{(\partial_{x} \lambda)^2}{\lambda-\lambda^2}+ \Lambda(x, y, \theta_{0}) (\partial_{x} h)^2,
\\
\\
F_{12}=F_{21}
\\
~~~~~~=\frac{(\partial_{x}\lambda) (\partial_{y} \lambda) }{\lambda-\lambda^2}+ \Lambda(x, y, \theta_{0}) (\partial_{x} h) (\partial_{y} h),
\\
\\
F_{22}=\frac{(\partial_{y} \lambda)^2}{\lambda-\lambda^2}+\Lambda(x, y, \theta_{0}) (\partial_{y} h)^2.
\end{array}
\end{equation}
It is clear that the determinant of the matrix $\textbf{F}_{xy}$ is given as
\begin{equation}
\begin{array}{llll}
\mathrm{Det}(\textbf{F}_{xy})=\frac{\Lambda(x, y, \theta_{0})}{\lambda-\lambda^2}( \partial_{x} \lambda \partial_{y} h - \partial_{y} \lambda \partial_{x} h)^2
\end{array}\label{detcon}
\end{equation}
with function $\Lambda(x, y, \theta_{0})=(3+\cos(2 \theta_{0})+2\cos(4h)\sin(\theta_{0})^{2})(1-2 \lambda)^2 \neq 0$
in general, and the factor in the parentheses determining whether $\mathrm{Det}(\textbf{F}_{xy})=0$.

Recalling that the density matrix of the qubit can be written as
$$\rho=\frac{1}{2}(\mathrm{I} + \overrightarrow{\mathrm{W}} \cdot \vec{\sigma})$$ with the Pauli matrix $\vec{{\sigma}}=(\sigma_{x}, \sigma_{y}, \sigma_{z})$. Therefore,
the two functions $\lambda(x,y)$ and $h(x,y)$ can be expressed in terms of $\mathrm{W}$ as
\begin{equation}
\begin{array}{llll}
\lambda(x, y)=(1-|\mathrm{W}|)/2,
\\
\\
h(x, y)=\arcsin[ \sqrt{ (|\mathrm{W}|+ w_{3})/(2 |\mathrm{W}|)}],
\end{array} \label{paraequ}
\end{equation}
with $\vec{\mathrm{W}}=(w_{1}, w_{2}, w_{3})$ being the real Bloch vector and $|\mathrm{W}|=\sqrt{w_{1}^2+w_{2}^2+w_{3}^2}$. Here $w_{i}\equiv w_{i}(x, y)$ are also the arbitary functions of $x$ and $y$. Substituting Eq.~(\ref{paraequ}) into Eq.~(\ref{detcon}), one can easily show that as long as the condition
\begin{equation}
\begin{array}{llll}
\partial_{x} |\mathrm{W}| \partial_{y} w_{3} - \partial_{y} |\mathrm{W}| \partial_{x} w_{3}= 0
\end{array} \label{revcondB}
\end{equation}
is satisfied, the inverse matrix of the QFIM $\textbf{F}$ does not exist. Therefore, the error of the joint estimation of the two parameters $x$ and $y$ is unbounded. Eq.~(\ref{revcondB}) is the main result of this paper.

\section{Example: a dissipative qubit}
As an application of the invertible condition Eq.~(\ref{revcondB}), we consider a dissipative qubit. It has the two levels $|e\rangle$ and $|g\rangle$, and its spontaneous emission rate of the excited state is $\gamma$. With the Born-Markov approximation, the time evolution of the system is ($\hbar=1$)
\begin{equation}
\begin{array}{lll}
\frac{d \rho}{d t}=-i[H, \rho]+ \gamma \mathcal{D}[\sigma_{-}]\rho.
\end{array} \label{liouv}
\end{equation}
Here the Hamiltonian is given as $H=\Omega \sigma_{z}$,
the superoperator is defined as
$\mathcal{D}[c]\rho \equiv c\rho c^{\dag}-\frac{1}{2}(c^{\dag} c \rho+ \rho c^{\dag} c )$, and
the Pauli operators $\sigma_{z}= |e\rangle \langle e|- |g\rangle \langle g|$. We set the time $t$ and the decay rate $\gamma$ are dimensionless parameters by scaling them with a time unit $t_{0}$.

For the initial state $\rho_{0}=x |e\rangle \langle e|+ (1-x) |g\rangle \langle g|$, the
state of the qubit at any time $t$ is solved as
\begin{equation}
\rho(t)=\left(                 
\begin{array}{cc}   
 \rho_{11}(t) & 0 \\  
 0 &  1-\rho_{11}(t) \\  
 \end{array}
 \right) \label{densityA}
\end{equation}
with the element $\rho_{11}(t)= e^{-t \gamma} x$.

As an example, we study the QFIM in terms of two parameters: the initial probability in the excited state
$x$ and the decay rate $\gamma$.
For the state in Eq.~(\ref{densityA}), it's easy to check that
the condition Eq.~(\ref{revcondB}) is satisfied. According to the multiple-parameter quantum Cram\'{e}r-Rao inequality, it is impossible to precisely estimate $x$ and $\gamma$ simultaneously.

Actually, it is easy to obtain its QFIM with respect to $x$ and $\gamma$
\begin{equation}
\textbf{F}_{\gamma,~x}=\left(                 
\begin{array}{cc}   
 F_{\gamma\gamma} & F_{\gamma x}
 \\
 \\  
 F_{x \gamma} &  F_{x x} \\  
 \end{array}
 \right)
\end{equation}
with the elements
\begin{equation}
\begin{array}{llll}
F_{\gamma\gamma}= t^2 x/(e^{t \gamma}-x),
~~F_{\gamma x}=F_{x\gamma}=t/(x-e^{t \gamma}),
\\
\\
F_{x x}=1/(xe^{t \gamma}-x^2).
\end{array}
\end{equation}
It is easy to check that
\begin{equation}
\begin{array}{llll}
\mathrm{Det}(\textbf{F}_{x,~\gamma})= 0,
\end{array}\label{FeedbackCond}
\end{equation}
which implies the inverse matrix of $\textbf{F}_{x,~\gamma}$ does not exist.

Although the parameters $x$ and $\gamma$ can not be estimated jointly, it is interesting to
study whether there are combination of the two parameters which can indeed be estimated.
We consider this question as following.
The QFIM $\textbf{F}_{\gamma,~x}$ can be transformed to the diagonal form
\begin{equation}
\textbf{F}_{\lambda_{1},~\lambda_{2}}=\left(                 
\begin{array}{cc}   
 F_{\lambda_{1}\lambda_{1}}  & 0
 \\
 \\  
 0 &  0\\  
 \end{array}
 \right)=\textbf{R} \textbf{F}_{\gamma,~x} \textbf{R}^{T}.
\end{equation}
Here the combined
new parameters are determined $[\lambda_{1},~\lambda_{2}]^{T}=\textbf{R}~[\gamma, x ]^{T}$
with the orthogonal matrix
\begin{equation}
\textbf{R}=\left(                 
\begin{array}{cc}   
  -tx & 1
 \\
 \\  
 1 &  tx
 \end{array}
 \right)/\sqrt{1 + (t x)^2}.
\end{equation}
\\
Thus, only the combined new parameter $\lambda_{1}=x(1 - t \gamma)/\sqrt{1 + t^2 x^2}$ can
be estimated, with the corresponding QFI $F_{\lambda_{1}\lambda_{1}}=(1 + t^2 x^2)/(e^{t \gamma}x - x^2)$.
The new parameter $\lambda_{2}$ can not be estimated completely.

\section{Conclusion}
Generally speaking, the property of a realistic system is determined by the multiple parameters.
It is expected that taking advantage of quantum metrology,
these parameters can be simultaneously estimated. According to the quantum Cram\'{e}r-Rao bound,
the lower bound on the total variance of all the estimated parameters
is related to the inverse matrix of QFIM. Based on the invertibility of the QFIM,
a condition, which determines whether two parameters
can be estimated simultaneously for a qubit in the mixed state, is given in this paper. We also give an example to show the application of this condition. In future, it's an interesting topic to explore the multi-qubits
to multiparameter estimation.
\\

\textit{Acknowledgments.}
This work is partially supported by the National Natural Science
Foundation of China (Grant Nos.~11365006,~11165006).




\begin{thebibliography}{99}
\bibitem{VGio11}
V. Giovanetti, S. Lloyd and L. Maccone, Science \textbf{306}, 1330
(2004).


\bibitem{Udem02}
N. Hinkley, J. A. Sherman, N. B. Phillips, M. Schioppo, N. D. Lemke, K. Beloy, M. Pizzocaro, C.W. Oates, and A. D. Ludlow, Science \textbf{341}, 1215 (2013).


\bibitem{Bollinger}
J. J. Bollinger, W. M. Itano, D. J. Wineland, and D. J. Heinzen,
Phys. Rev. A \textbf{54}, R4649 (1996).



\bibitem{Aasi2}
J. Aasi et al. (LIGO Scientific Collaboration), Nat. Photon. \textbf{7},
613 (2013); U. L. Andersen, Nat. Photon. \textbf{7}, 589 (2013).


\bibitem{Taylor08}
J. M. Taylor, P. Cappellaro, L. Childress, L. Jiang, D. Budker,
P. R. Hemmer, A. Yacoby, R. Walsworth, and M. D. Lukin,
Nat. Phys. \textbf{4}, 810 (2008).


\bibitem{Escher11}
B. M. Escher, R. L. de Matos Filho, and L. Davidovich, Nat.
Phys. \textbf{7}, 406 (2011).


\bibitem{Guta12}
R. Demkowicz-Dobrza\'{n}ski, J. Kolody\'{n}ski, and M. Gut\u{a},
Nat. Commun. \textbf{3}, 1063 (2012).


\bibitem{zyou14}
F. Hudelist, J. Kong, C. Liu, J. Jing, Z. Y. Ou, and W. P. Zhang,
Nat. Commun. \textbf{5}, 3049 (2014).


\bibitem{jin13}
Y. M. Zhang, X. W. Li, W. Yang, and G. R. Jin, Phys.
Rev. A \textbf{88}, 043832 (2013).



\bibitem{qz14}
Q. Zheng, Y. Yao and Y. Li, Eur. Phys. J. D
\textbf{68}, 170 (2014).



\bibitem{wzhong13}
W. Zhong, Z. Sun, J. Ma, X. G. Wang, and F. Nori,
Phys. Rev. A \textbf{87}, 022337 (2013).


\bibitem{Smerzi09}
L. Pezz$\acute{e}$ and A. Smerzi, Phys. Rev. Lett. \textbf{102}, 100401 (2009);
N. Li and S. L. Luo, Phys. Rev. A \textbf{88}, 014301 (2013).



\bibitem{xmLu10}
X. M. Lu, X. G. Wang, and C. P. Sun, Phys. Rev. A \textbf{82}, 042103
(2010).


\bibitem{jma11}
J. Ma, X. G. Wang, C. P. Sun, and F. Nori, Phys. Rep.
\textbf{509}, 89 (2011);
X. G. Wang, A. Miranowicz, Y. X. Liu, C. P. Sun, and F. Nori, Phys. Rev. A
\textbf{81}, 022106 (2010).




\bibitem{yuen73}
H. P. Yuen and M. Lax, IEEE Trans. Inf. Theory \textbf{19}, 740 (1973).

\bibitem{Helstrom74}
C. Helstrom and R. Kennedy, IEEE Trans. Inf. Theory \textbf{20}, 16
(1974).



\bibitem{Ballester04}
M. A. Ballester, Phys. Rev. A \textbf{69}, 022303 (2004);
M. A. Ballester, Phys. Rev. A \textbf{70}, 032310 (2004).



\bibitem{Walmsley14}
M. D. Vidrighin, G. Donati, M. G. Genoni, X. M. Jin,
W. S. Kolthammer, M. S. Kim, A. Datta, M. Barbieri, and
I. A. Walmsley, Nat. Commun. \textbf{5}, 3532 (2014).




\bibitem{Vijay12}
R. Vijay, C. Macklin, D. H. Slichter,
S. J. Weber, K. W. Murch, R. Naik, A. N. Korotkov, and I. Siddiqi,
Nature \textbf{490}, 77 (2012).



\bibitem{Holevo} A. S. Holevo,
\textit{Probabilistic and Statistical Aspects of Quantum Theory}
(North-Holland, Amsterdam, 1982).


\bibitem{Fisher}
R. A. Fisher, Proc. Cambridge Philos. Soc. \textbf{22}, 700
(1925).


\bibitem{paris09}
M. G. A. Paris, Int. J. Quant. Inf. \textbf{7}, 125 (2009).


\bibitem{Caves94}
S. L. Braunstein and C. M. Caves, Phys. Rev. Lett.
\textbf{72}, 3439 (1994).


\bibitem{Caves96}
S. L. Braunstein, C. M. Caves, and G. J. Milburn, Ann.
Phys. (N.Y.) \textbf{247}, 135 (1996).



\bibitem{Nielsen00}
M. A. Nielsen and I. L. Chuang, \textit{Quantum Computation
and Quantum Information} (Cambridge University Press,
Cambridge, U.K., 2000).



\bibitem{Genoni08}
M. G. Genoni, P. Giorda, and M. G. A. Paris, Phys. Rev. A \textbf{78},
032303 (2008).


\bibitem{Whaley09}
K. C. Young, M. Sarovar, R. Kosut, and K. B. Whaley, Phys.
Rev. A \textbf{79}, 062301 (2009).



\bibitem{Monras10}
A. Monras and F. Illuminati, Phys. Rev. A \textbf{83}, 012315
(2011).


\bibitem{Kim13}
M. G. Genoni, M. G. A. Paris, G. Adesso, H. Nha,
P. L. Knight, and M. S. Kim, Phys. Rev. A \textbf{87}, 012107
(2013).



\bibitem{Fujiwara01}
A. Fujiwara, Phys. Rev. A 65, 012316 (2001); K. Matsumoto, J. Phys. A 35, 3111 (2002);
M. Gu\c{t}\v{a} andA. Jen\v{c}ov\'{a}, Commun. Math. Phys. 276, 341 (2007);
J. Kahn and M. Gu\c{t}\v{a}, Commun. Math. Phys. 289, 597 (2009).



\bibitem{Walmsley13}
P. C. Humphreys, M. Barbieri, A. Datta, and I. A. Walmsley,
Phys. Rev. Lett. \textbf{111}, 070403 (2013).


\bibitem{hfan14}
J. D. Yue, Y. R. Zhang, and H. Fan, Sci. Rep. \textbf{4}, 5933 (2014).






\end{thebibliography}
\end{document}